\documentclass[11pt,aps,onecolumn,nopacs,nofootinbib,floatfix,superscriptaddress]{revtex4}
\usepackage[utf8]{inputenc}
\usepackage{graphicx}
\usepackage{amsfonts}
\usepackage{amssymb}
\usepackage{amsbsy}
\usepackage{hyperref}
\usepackage{amsmath}
\usepackage{mathrsfs}
\usepackage{latexsym}
\usepackage{natbib}
\usepackage{bm}
\usepackage{subfigure} 
\usepackage{color}
\usepackage{wasysym}
\usepackage{mathbbol}
\usepackage{bigints}
\allowdisplaybreaks
\usepackage[normalem]{ulem}
\usepackage[dvipsnames]{xcolor}
\usepackage{graphicx}
\usepackage{subcaption} 
\usepackage{amsfonts}
\usepackage{amssymb}
\usepackage{amsbsy}
\usepackage{hyperref}
\usepackage{amsmath}
\usepackage{mathrsfs}
\usepackage{latexsym}
\usepackage{natbib}
\usepackage{bm}
\usepackage{physics}
\usepackage{subfigure} 
\usepackage{color}
\usepackage{wasysym}
\usepackage{mathbbol}
\usepackage{bigints}
\usepackage{comment}
\allowdisplaybreaks
\usepackage[normalem]{ulem}
\usepackage[dvipsnames]{xcolor}
\usepackage{multirow}
\usepackage{csquotes}
\usepackage{hyperref}
\usepackage{graphicx}
\usepackage{float} 
\usepackage{tikz}
\usepackage{pgfplots}
\pgfplotsset{compat=1.18}
\usepackage{orcidlink}
\usepackage{tikz-3dplot}
\usepackage{float}
\usepackage{subcaption}
\usepackage[paperwidth=210mm,paperheight=297mm,centering,hmargin=2cm,vmargin=2.6cm]{geometry}
\usepackage{tikz}
\usetikzlibrary{arrows.meta,calc}
\definecolor{napiergreen}{rgb}{0.16, 0.5, 0.0}

\begin{document}


\title{\Large{Memory-like effects and kinematics of trajectories \\in Cyclotron motion}}


\author{Manthan Kashyap Datta\,\orcidlink{0009-0007-8501-6288}}
\email{manthandatta05@gmail.com}
\affiliation{Department of Physics, Birla Institute of Technology and Science - Pilani, Rajasthan, 333031, India}

\author{Mantra Mehta\, \orcidlink{0009-0007-0498-9660}}
\email{mantramehta8364@gmail.com}
\affiliation{Department of Physics, Birla Institute of Technology and Science - Pilani, Rajasthan, 333031, India}
\author{Sayan Das\,\orcidlink{0009-0000-2420-767X}}
\email{p20240080@pilani.bits-pilani.ac.in}
\affiliation{Department of Physics, Birla Institute of Technology and Science - Pilani, Rajasthan, 333031, India}

\begin{abstract}
 We investigate the collective dynamics of a bundle of charged particles undergoing cyclotron motion in a uniform magnetic field when subjected to a short-duration electric pulse. Using the geometric framework based on the evolution of trajectory congruences, we analyze how the pulse affects the expansion, shear, and rotation of a small family of trajectories. We show that the geometric imprint persists after the pulse has vanished, manifesting as a \textit{memory} of the transient perturbation. Unlike gravitational memory effects, this does not manifest itself in focusing behaviour of the trajectories, and instead implies a restructuring of the shear component before and after the pulse. We offer direct analytic and regression based arguments for the same. 
\end{abstract}

\maketitle

\newpage
\section{Introduction}In recent years, memory effects have emerged as an important probe of the interplay between geometry and dynamics in a variety of physical systems. Broadly speaking, a memory effect refers to a permanent change in the state or configuration of a system induced by a passage of a transient disturbance. Among the most prominent examples is the gravitational-wave memory effect, originally discussed by Zel'dovich , Polnarev \cite{Zeldovich:1974gvh} and Braginsky, Grishchuk \cite{Braginsky:1985vlg} and later developed through the pioneering work of Christodoulou \cite{Christodoulou:1991cr}. Since then, gravitational-wave memory has been studied extensively, leading to a rich body of work that has deepened our understanding of the phenomenon and its physical implications \cite{PhysRevD1,PhysRevD2,Tolish:2014bka,Madler:2016ggp,Madler:2017umy,Strominger:2014pwa,Strominger:2017zoo,Zhang:2017-1,Zhang:2017-2,Zhang:2017-3,Zhang:2018srn,Zhang:2018gzn:2018-1,Shore:2018kmt:2018-2,Flanagan:2019ezo,OLoughlin:2018ebk,Chakraborty:2019yxn,Chakraborty:2020uui,Siddhant:2020gkn,Divakarla:2021xrd,chakraborty2022simple,Cvetkovic:2021aqh,Grant:2021hga,Zhang:2024-1,BenAchour:2024ucn,Hadi:2024idb,Zhang:2024-2, Hait:2022ukn,9gqz-sl9n}. Comprehensive discussions of gravitational-wave memory and its subsequent developments can be found in \cite{Favata:2010zu,Mitman:2024uss, Mohanty:2022abo}\footnote{This list is in no way exhaustive, please see further references.}.

Memory phenomena are not only confined to gravitational systems and have
also been investigated in electromagnetic settings \cite{hjrp-5cps}.
\textcolor{black}{In analogy with gravitational wave memory, electromagnetic memory is
conventionally defined also through a ``permanent change''---the
so-called ``velocity kick''. It was first briefly discussed in a paper
by Grishchuk and Polnarev \cite{Grishchuk:1989qa} and has recently been extensively
analysed by Bieri and Garfinkle \cite{Bieri:2013hqa}}. Subsequent studies have examined various aspects of electromagnetic memory, including its global \cite{Winicour:2014ska}
and asymptotic properties \cite{Pasterski:2015zua}, duality structures \cite{Hamada:2017bgi}, connection
with wave propagation \cite{Garfinkle:2022dnm} and potential experimental manifestations
\cite{Bieri:2023btq}.

Despite arising in a variety of physical systems, many memory phenomena share a common geometric feature. They are ultimately manifested through the relative evolution of neighboring probes under the influence of a transient disturbance. In gravitational systems, memory manifests through lasting changes in the relative motion of freely falling test particles, naturally motivating the study of geodesic congruences. The evolution of a geodesic congruence is characterized by three kinematical quantities: expansion, which describes the overall convergence or divergence of neighboring geodesics; shear, which measures distortions in the shape of the congruence without changing its volume; and rotation, which characterizes the local twisting of the trajectory bundle \cite{PhysRev.98.1123,Hawking:1973uf}. The dynamics of these quantities (ESR henceforth) are governed by the Raychaudhuri equation \footnote{For a beautiful review of the Raychaudhuri Equation see \cite{Kar:2006ms}.}, a fundamental result that relates the evolution of a congruence to the geometry of spacetime \cite{PhysRev.98.1123}. Owing to its central role in the study of gravitational focusing and the Hawking--Penrose singularity theorems \cite{Hawking:1973uf,Penrose:1964wq,Hawking:1970zqf}, the Raychaudhuri equation provides a powerful geometric framework for understanding the relative motion of test particles and the emergence of memory effects.

Guided by the geometric perspective provided by congruence dynamics, the analysis in the current work employs memory effects within a trajectory-based ESR framework for classical systems, in the same lines as in \cite{shaikh2014kinematics}. The versatility of the ESR formalism has been demonstrated in a variety of gravitational settings, including studies of geodesic congruences in gravitational collapse, where the evolution of expansion, shear, and rotation has been used to characterize focusing, geodesic deviation, and spacetime dynamics \cite{Shaikh:2014xna}. By tracking the evolution of neighboring trajectories through ESR variables, this approach offers a natural means of characterizing the imprints left by transient perturbations, focusing on a very particular system: motion in a cyclotron. This intriguing example was noted in \cite{kar2024pulse}, wherein a short-duration electric pulse, switched on in the background of uniform cyclotron motion, was shown to induce a permanent change in the orbit's center, radius, and velocity. This is dubbed as a \textit{memory-like} observable effect. Our current analysis probes this physics by examining how such a pulse affects the collective behavior of a congruence of trajectories. We do this by explicitly computing the ESR variables in the temporal regions separated by the pulse with separate assumptions on the focusing times. Unlike the cases of gravitational pulses passing through \cite{Harte:2024mwj}, the focusing time here does not explicitly depend on the width and duration of the eletric field pulse. This is a subtle situation, as the focusing/defocusing of the congruence does not encode the memory of the pulse. Our analysis finds that from a congruence perspective, the memory is encoded into the shear variable, which seamlessly connects to the observation in \cite{kar2024pulse} that the pulse induces a gauge transformation on the system. 

The rest of the paper is arranged as following. 
Section \eqref{section-II} presents a review of the ESR formalism for two-dimensional classical systems and the associated congruence dynamics. Here we study the evolution of ESR variables for a pulse-driven charged particle in a uniform magnetic field and analyze the corresponding focusing behaviour. Section \eqref{section-IV} focuses on the shear sector, where we demonstrate the persistence of memory-like signatures and validate the analytical predictions through a regression-based analysis, together with a discussion of their possible experimental relevance. Finally, Section \eqref{section-V} explores the role of complex shear and rotation variables, their connection to electromagnetic phase memory and gauge transformations, and their interpretation in terms of a dynamical phase. In \eqref{section-VI} we conclude with a summary and plans for future extensions.

\section{GEODESIC CONGRUENCE IN 2D FOR CLASSICAL MECHANICS}\label{section-II}

The dynamics of an isolated trajectory generally provides only limited information about the underlying structure of a dynamical system. A much richer characterization emerges from the study of a \textit{congruence} or a family of nearby trajectories generated by infinitesimal perturbations of the initial conditions. The evolution of such a family probes the local geometry of the configuration space and reveals collective features that remain invisible at the level of individual trajectories. This perspective is closely analogous to the study of geodesic congruences in general relativity \cite{Poisson:2009pwt}, where the relative motion of neighboring geodesics encodes important geometric and physical information about the spacetime itself. For interesting applications of geodesic congruence, see for example \cite{Dasgupta:2007nr,Dasgupta:2008in,Dasgupta:2008jt,Ghosh:2010gq,Dasgupta:2012zf}.

The central idea is to quantify how initially nearby trajectories evolve with respect to one another. As the system evolves, the congruence may undergo local expansion or contraction, shear-induced distortions that alter its shape while preserving its area, and rotational motion corresponding to local vorticity. These kinematical characteristics provide a geometric description of the flow and furnish valuable information regarding stability, focusing behaviour, and the emergence of singular structures within the dynamical evolution. The formalism developed in \cite{shaikh2014kinematics}, especially for classical dynamics, offers a systematic framework for analyzing these effects through the ESR variables, thereby enabling a detailed characterization of the evolution of trajectory congruences in a broad class of dynamical systems.

\subsection{ESR Variables}

Let us now provide some technical details of our setup.
Consider a velocity field $u^i(x,t)$ defined in an $n$-dimensional configuration space. The relative motion of nearby trajectories for such a system is governed by the spatial variation of this velocity field. This information is encoded in the velocity gradient tensor:
\begin{equation}
B^i_{\; j} = \partial_j u^i.
\label{eq:B_def}
\end{equation}

To determine how this tensor evolves along the trajectory flow, we take the convective derivative. The result is given by \cite{shaikh2014kinematics}:

\begin{equation}
u^k \partial_k B^i_{\; j} = \partial_j f^i - B^i_{\; k} B^k_{\; j}.
\label{eq:B_evolution}
\end{equation}

Equation~\eqref{eq:B_evolution} serves as the fundamental evolution equation for the kinematics of a trajectory congruence. The tensor $B^i_{\; j}$ can be decomposed uniquely into its irreducible components under the action of the orthogonal group: an isotropic expansion, a traceless symmetric shear, and an antisymmetric rotation. In $n$ dimensions, this decomposition reads:
\begin{equation}
B^i_{\; j} = \frac{1}{n} \theta \, \delta^i_{\; j} + \sigma^i_{\; j} + \omega^i_{\; j},
\label{eq:B_decomposition}
\end{equation}
where the expansion scalar $\theta$, the shear tensor $\sigma_{ij}$, and the rotation tensor $\omega_{ij}$ are defined via the covariant components $B_{ij} = g_{ik} B^k_{\; j}$ (with $g_{ij}$ the Euclidean metric) as:
\begin{align}
\theta &= B^i_{\; i}, \label{eq:theta_def} \\
\sigma_{ij} &= \frac{1}{2}\bigl(B_{ij} + B_{ji}\bigr) - \frac{1}{n}\,\delta_{ij}\,\theta, \label{eq:shear_def} \\
\omega_{ij} &= \frac{1}{2}\bigl(B_{ij} - B_{ji}\bigr). \label{eq:rotation_def}
\end{align}
Equivalently, the mixed components are $\sigma^i_{\; j} = g^{ik}\sigma_{kj}$ and $\omega^i_{\; j} = g^{ik}\omega_{kj}$, both of which are traceless and satisfy $\sigma^i_{\; i}=0$, $\omega^i_{\; i}=0$.

Substituting the decomposition \eqref{eq:B_decomposition} into the evolution equation \eqref{eq:B_evolution} yields a coupled system of ordinary differential equations for $\theta$, $\sigma_{ij}$, and $\omega_{ij}$ along the flow. These are the Raychaudhuri equations for a classical velocity field.

\subsection{Specialization to Two Dimensions}

In two dimensions ($n = 2$), the general decomposition of $B^i_{\; j}$ simplifies while retaining all three kinematic effects. 

Inserting this decomposition into the general evolution equation \eqref{eq:B_evolution} yields the following system of ordinary differential equations for the kinematic variables along the flow:
\begin{align}
\frac{d\theta}{dt} + \frac{1}{2}\theta^2 + 2\left(\sigma_+^2 + \sigma_\times^2 - \omega^2\right)
&= \frac{\partial f_x}{\partial x} + \frac{\partial f_y}{\partial y},
\label{eq:theta_evolution} \\
\frac{d\sigma_+}{dt} + \theta \sigma_+ 
&= \frac{1}{2}\left(\frac{\partial f_x}{\partial x} - \frac{\partial f_y}{\partial y}\right),
\label{eq:sigmap_evolution} \\
\frac{d\sigma_\times}{dt} + \theta \sigma_\times 
&= \frac{1}{2}\left(\frac{\partial f_x}{\partial y} + \frac{\partial f_y}{\partial x}\right),
\label{eq:sigmax_evolution} \\
\frac{d\omega}{dt} + \theta \omega 
&= \frac{1}{2}\left(\frac{\partial f_x}{\partial y} - \frac{\partial f_y}{\partial x}\right).
\label{eq:omega_evolution}
\end{align}
Here $f_x$ and $f_y$ are the components of the force per unit mass (acceleration field), and the derivatives on the right-hand side are evaluated along the central trajectory in the bunch of geodesics.

These equations govern the complete time evolution of a bundle of trajectories in two dimensions. A key consequence is that the occurrence of trajectory focusing — i.e., the intersection of nearby trajectories — is determined solely by the initial values of the ESR variables $(\theta_0, \sigma_{+0}, \sigma_{\times 0}, \omega_0)$ and the right-hand side source terms generated by the force fields. In particular, for force fields that satisfy $\partial_j f^i = \partial_i f^j$ (i.e., for a potential force), the evolution of $\omega$ decouples and focusing can be analyzed via the Raychaudhuri equation for $\theta$ only.

\subsection{Geometric Interpretation via a Square Configuration}
To visualize the physical meaning of the kinematic variables, we will provide a similar discussion as in \cite{shaikh2014kinematics}. We consider a small square centered at $(x_0, y_0)$ in the $xy$-plane. Four trajectories start from its corners, with the central trajectory passing through the center. Let $(u_{x0}, u_{y0})$ be the velocity of the central trajectory at the initial time. For sufficiently small initial separations, the velocity field at the corners can be approximated by a first-order Taylor expansion:
\begin{align}
u_{xi0} &= u_x(x_0,y_0) + \frac{\partial u_x}{\partial x}\bigg|_0 \Delta x_{i0}
+ \frac{\partial u_x}{\partial y}\bigg|_0 \Delta y_{i0},
\label{eq:taylor_x} \\
u_{yi0} &= u_y(x_0,y_0) + \frac{\partial u_x}{\partial x}\bigg|_0 \Delta x_{i0}
+ \frac{\partial u_y}{\partial y}\bigg|_0 \Delta y_{i0},
\label{eq:taylor_y}
\end{align}
where $(\Delta x_{i0}, \Delta y_{i0})$ are the initial offsets of the $i$-th corner relative to the center (see Figure. \eqref{fig:initial_configuration}). The partial derivatives are evaluated at the center at $t=0$.

Using the ESR decomposition of the velocity gradient tensor $B^i_{\; j} = \partial_j u^i$ and substituting the matrix elements into the Taylor expansion yields velovities at the corners as:
\begin{equation}
\begin{pmatrix}
u_{xi0} \\
u_{yi0}
\end{pmatrix}
=
\begin{pmatrix}
u_{x0} \\
u_{y0}
\end{pmatrix}
+
\begin{pmatrix}
\frac{\theta_0}{2} + \sigma_{+0} & \sigma_{\times 0} + \omega_0 \\
\sigma_{\times 0} - \omega_0 & \frac{\theta_0}{2} - \sigma_{+0}
\end{pmatrix}
\begin{pmatrix}
\Delta x_{i0} \\
\Delta y_{i0}
\end{pmatrix},
\label{eq:corner_velocity_final}
\end{equation}
where the subscript $0$ denotes initial-time values of the ESR variables.

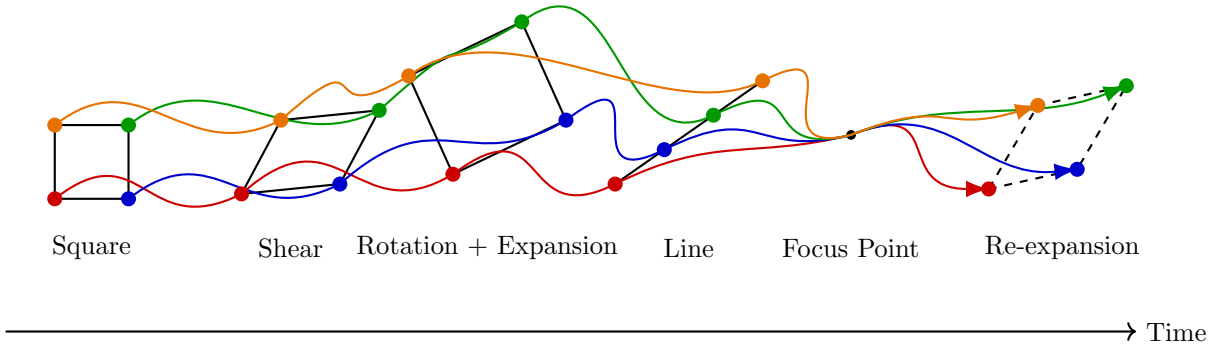
\begin{figure}[htbp]
    \centering
\begin{tikzpicture}[scale=0.65,
    particle/.style={circle,inner sep=2pt},
    frame/.style={thick,black},
    every node/.style={font=\small}
]

\def\tA{0}
\def\tB{3.8}
\def\tC{7.6}
\def\tD{11.4}
\def\tE{15.2}
\def\tF{19}

\coordinate (A1) at (\tA,0);
\coordinate (B1) at (\tA+1.5,0);
\coordinate (C1) at (\tA+1.5,1.5);
\coordinate (D1) at (\tA,1.5);

\draw[frame] (A1)--(B1)--(C1)--(D1)--cycle;
\node at (\tA+0.75,-1.0) {Square};

\coordinate (A2) at (\tB,0.1);
\coordinate (B2) at (\tB+2.0,0.3);
\coordinate (C2) at (\tB+2.8,1.8);
\coordinate (D2) at (\tB+0.8,1.6);

\draw[frame] (A2)--(B2)--(C2)--(D2)--cycle;
\node at (\tB+1.0,-1.0) {Shear};

\coordinate (A3) at (\tC+0.5,0.5);
\coordinate (B3) at (\tC+2.8,1.6);
\coordinate (C3) at (\tC+1.9,3.6);
\coordinate (D3) at (\tC-0.4,2.5);

\draw[frame] (A3)--(B3)--(C3)--(D3)--cycle;
\node at (\tC+1.2,-1.0) {Rotation + Expansion};

\coordinate (A4) at (\tD,0.3);
\coordinate (B4) at (\tD+1.0,1.0);
\coordinate (C4) at (\tD+2.0,1.7);
\coordinate (D4) at (\tD+3.0,2.4);

\draw[frame] (A4)--(B4)--(C4)--(D4);
\node at (\tD+1.5,-1.0) {Line};

\coordinate (P) at (\tE+1,1.3);

\filldraw[black] (P) circle (2.5pt);
\node at (\tE+1,-1.0) {Focus Point};

\coordinate (A5) at (\tF,0.2);
\coordinate (B5) at (\tF+1.8,0.6);
\coordinate (C5) at (\tF+2.8,2.3);
\coordinate (D5) at (\tF+1.0,1.9);

\draw[frame,dashed] (A5)--(B5)--(C5)--(D5)--cycle;
\node at (\tF+1.5,-1.0) {Re-expansion};


\draw[thick,red!80!black,-{Latex[length=3mm]}]
(A1)
.. controls +(1.8,1.4) and +(-2.0,-1.0) .. (A2)
.. controls +(2.0,1.8) and +(-2.0,-1.2) .. (A3)
.. controls +(2.0,1.5) and +(-2.0,-1.0) .. (A4)
.. controls +(2.0,1.0) and +(-2.0,-0.5) .. (P)
.. controls +(2.0,0.8) and +(-2.0,0.0) .. (A5);

\draw[thick, blue!80!black,-{Latex[length=3mm]}]
(B1)
.. controls +(1.8,1.4) and +(-2.0,-1.0) .. (B2)
.. controls +(2.0,1.8) and +(-2.0,-1.2) .. (B3)
.. controls +(2.0,1.5) and +(-2.0,-1.0) .. (B4)
.. controls +(2.0,1.0) and +(-2.0,-0.5) .. (P)
.. controls +(2.0,0.8) and +(-2.0,-0.2) .. (B5);

\draw[thick, green!60!black,-{Latex[length=3mm]}]
(C1)
.. controls +(1.8,1.4) and +(-2.0,-1.0) .. (C2)
.. controls +(2.0,1.8) and +(-2.0,-1.2) .. (C3)
.. controls +(2.0,1.5) and +(-2.0,-1.0) .. (C4)
.. controls +(2.0,1.0) and +(-2.0,-0.5) .. (P)
.. controls +(2.0,0.8) and +(-2.0,-0.8) .. (C5);

\draw[thick, orange!90!black,-{Latex[length=3mm]}]
(D1)
.. controls +(1.8,1.4) and +(-2.0,-1.0) .. (D2)
.. controls +(2.0,1.8) and +(-2.0,-1.2) .. (D3)
.. controls +(2.0,1.5) and +(-2.0,-1.0) .. (D4)
.. controls +(2.0,1.0) and +(-2.0,-0.5) .. (P)
.. controls +(2.0,0.8) and +(-2.0,-0.6) .. (D5);

\foreach \p in {A1,A2,A3,A4,A5}
    \node[particle,fill=red!80!black] at (\p) {};

\foreach \p in {B1,B2,B3,B4,B5}
    \node[particle,fill=blue!80!black] at (\p) {};

\foreach \p in {C1,C2,C3,C4,C5}
    \node[particle,fill=green!60!black] at (\p) {};

\foreach \p in {D1,D2,D3,D4,D5}
    \node[particle,fill=orange!90!black] at (\p) {};

\draw[->,thick] (-1,-2.7) -- (22,-2.7) node[right] {Time };

\end{tikzpicture}
\caption{Schematic for evolution of a four-particle trajectory undergoing shear, rotation, expansion, collapse to a line and point, followed by re-expansion.}
    \label{fig:particle_deformation}
\end{figure}

The subsequent time evolution of the four corner trajectories, obtained by integrating their equations of motion, leads to a continuous deformation of the square. This deformation encodes the physical meaning of the kinematic variables:
\begin{itemize}
    \item $\theta$ (expansion) controls the overall change in area of the square.
    \item $\sigma_+$ and $\sigma_\times$ (shear) are responsible for shape distortion (e.g., stretching along the axes or diagonals) without altering the area.
    \item $\omega$ (rotation) causes the square to rotate as a whole about the central trajectory.
\end{itemize}
Thus, the ESR formalism provides a complete geometric description of the relative motion of nearby trajectories. See Figure. \eqref{fig:particle_deformation} for a visual description of the action of these kinematic variables as the trajectories move in time.

A more formal description of the relative motion is provided by \cite{shaikh2014kinematics}, using the deviation vector $\eta^i(t)$, but to our current analysis it is not of much importance.

This completes the formal kinematic framework for analyzing the relative motion of trajectory congruences in two-dimensional classical systems. We will now focus on a very particular situation, which will constitute the problem at hand.

\subsection{Kinematics of trajectories for particles in EM field}\label{subsection-D}

In this subsection, we analyze the collective behavior of trajectories for a charged particle of mass $m$ and charge $q$ in a uniform magnetic field $\mathbf{B} = B \hat{\mathbf{k}}$ and a short-duration electric field pulse. Initially, the particle follows a standard circular cyclotron path. We introduce a spatially homogeneous electric field pulse $\mathbf{E}(t)$ confined to the $xy$-plane, defined as:
\begin{equation}
\mathbf{E}(t) = E_0 \bigl( \cos\alpha\,\hat{\mathbf{i}} + \sin\alpha\,\hat{\mathbf{j}} \bigr) \, \Theta(t) \Theta(T-t),
\label{eq:E_field}
\end{equation}
where $E_0$ is the constant field magnitude, $T$ is the pulse duration, and $\Theta(\cdot)$ is the Heaviside step function. Equivalently,
\begin{equation}
\mathbf{E}(t) = \begin{cases}
\mathbf{0}, & t \le 0, \\[4pt]
E_0 (\cos\alpha\,\hat{\mathbf{i}} + \sin\alpha\,\hat{\mathbf{j}}), & 0 < t \le T, \\[4pt]
\mathbf{0}, & t > T.
\end{cases}
\label{eq:E_piecewise}
\end{equation}

\begin{figure}[htbp]
    \centering
    \begin{tikzpicture}[scale=1.0,
    axis/.style={thick,->},
    pulse/.style={very thick,blue},
    fieldlabel/.style={font=\small},
    dashedmark/.style={dashed,gray}
]

\def\Ezero{3}       
\def\tmin{-3}       
\def\T{6}           

\draw[axis] (\tmin,0) -- (9,0) node[right] {$t$};
\draw[axis] (\tmin,-0.5) -- (\tmin,4.2) node[above] {$E(t)$};

\draw[pulse]
(\tmin,0) -- (0,0)
(0,0) -- (0,\Ezero)
(0,\Ezero) -- (\T,\Ezero)
(\T,\Ezero) -- (\T,0)
(\T,0) -- (8.5,0);

\draw[dashedmark] (0,\Ezero) -- (\tmin,\Ezero);
\draw[dashedmark] (\T,0) -- (\T,-0.3);

\node[below] at (\tmin,-0.5) {$t_i$};
\node[below] at (0,0) {$0$};
\node[below] at (\T,0) {$T$};
\node[left] at (\tmin,\Ezero) {$E_0$};

\node[fieldlabel] at (6.8,3.7) {Background magnetic field: $\mathbf{B}=B\hat{k}$};

\draw[thick] (7.8,2.8) circle (0.25);
\fill (7.8,2.8) circle (0.06);
\node[right] at (8.1,2.8) {$B\hat{k}$};


\end{tikzpicture}

    \caption{Schematic of the pulse. Electric field is zero for $t<t_i$, turns on at $t=0$, and switches off at $t=T$.}
    \label{fig:Pulse}
\end{figure}
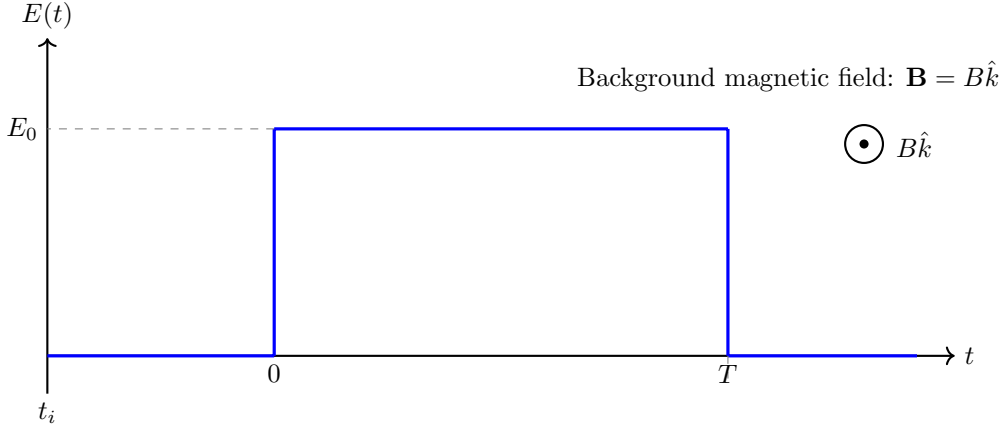

This pulse defines three distinct temporal regions:
\begin{itemize}
    \item \textbf{Region I} ($t \le 0$): No electric field,
    \item \textbf{Region II} ($0 < t \le T$): Constant electric field pulse,
    \item \textbf{Region III} ($t > T$): Electric field vanishes again.
\end{itemize}
See Figure. \eqref{fig:Pulse} for a visualization.
In Regions I and III, the particle moves on a circular cyclotron orbit (pure magnetic field). However, the Electric pulse switched on for finite time in Region II induces a \emph{permanent} change in the orbit's center, radius, and velocity; a phenomenon identified as an electromagnetic memory-like effect~\cite{kar2024pulse}. The equations of motion during the pulse (Region II) are obtained from the Lorentz force law:
\begin{align}
m \frac{d v_x}{dt} &= q E_0 \cos\alpha + q B\, v_y, \label{eq:mot_x} \\
m \frac{d v_y}{dt} &= q E_0 \sin\alpha - q B\, v_x, \label{eq:mot_y} \\
m \frac{d v_z}{dt} &= 0. \label{eq:mot_z}
\end{align}
Since the magnetic field is uniform and the electric field has no $z$-component, the motion remains confined to the $xy$-plane if the initial $v_z$ is zero. The primary objective, however, is not only the detailed solution of these equations, but rather the evolution of the expansion, shear, and rotation (ESR) variables across the discontinuous jump of the electric field. 

\subsection{Equations of Motion and ESR Formalism for the three regions}

In Regions I and III, the electric field evidently vanishes. The ESR variables in this case satisfy the following system (derived from the velocity gradient tensor and the Raychaudhuri equation):
\begin{align}
\frac{d\theta}{dt} + \frac{1}{2}\theta^2 + 2\bigl(\sigma_+^2 + \sigma_\times^2 - \overline{\omega}^2\bigr) + \frac{\omega_c^2}{2} &= 0, \label{eq:theta_ev}\\
\frac{d\sigma_+}{dt} + \theta\sigma_+ - \omega_c\sigma_\times &= 0, \label{eq:sigmap_ev}\\
\frac{d\sigma_\times}{dt} + \theta\sigma_\times + \omega_c\sigma_+ &= 0, \label{eq:sigmax_ev}\\
\frac{d\overline{\omega}}{dt} + \theta\overline{\omega} &= 0, \label{eq:omegab_ev}
\end{align}
where $\omega_c = \frac{qB}{m}$ is the cyclotron frequency and we have defined $\overline{\omega} = \omega - \frac{\omega_c}{2}$. The quantity $I = \sigma_+^2 + \sigma_\times^2 - \overline{\omega}^2$ is an invariant in the absence of external forces, and this determines the solution space, which of course depend on the sign of $I$. Actually this determines whether a congruence of trajectories will focus (converge to a point) at a finite time. This invariant emerges naturally from the evolution equations and serves as a discriminant for the qualitative behavior of the trajectory bundle. On a physical sense \(\sigma_+^2 + \sigma_\times^2\) measures the magnitude of shear deformation  (area-preserving distortion) and \(\overline{\omega}^2\) measures the effective rotation of the bundle, with the quantity \(I\) representing the balance between these two. 
The focusing condition corresponds to the divergence of the expansion scalar \(\theta(t)\), which occurs when the denominator of the solutions vanishes at a finite time \(t_f\).
\subsubsection{Case 1: $I > 0$}

The solutions for ESR variables in this case reads:
\begin{align}
\theta(t) &= \frac{\omega_c\sqrt{1+C^2}\cos(\omega_c t + D)}{C + \sqrt{1+C^2}\,\sin(\omega_c t + D)}, \notag\\
\sigma_+(t) &= \frac{E\sin(\omega_c t) + F\cos(\omega_c t)}{C + \sqrt{1+C^2}\,\sin(\omega_c t + D)}, \notag\\
\sigma_\times(t) &= \frac{E\cos(\omega_c t) - F\sin(\omega_c t)}{C + \sqrt{1+C^2}\,\sin(\omega_c t + D)}, \notag\\
\omega(t) &= \frac{\omega_c}{2} + \frac{G}{C + \sqrt{1+C^2}\,\sin(\omega_c t + D)},
\label{equ-31}
\end{align}
with constants depending on initial values:
\begin{align}
C &= \frac{\sqrt{I_0}}{\omega_c}\left(-1 + \frac{\theta_0^2}{4I_0} + \frac{\omega_c^2}{4I_0}\right), &
D &= \tan^{-1}\!\left[\frac{2\sqrt{I_0}}{\theta_0}\left(\frac{\omega_c}{2\sqrt{I_0}} - C\right)\right],\\
\{E,F,G\} &= \frac{\omega_c}{2\sqrt{I_0}}\left\{\sigma_{+0},\;\sigma_{\times0},\;\omega_0 - \frac{\omega_c}{2}\right\}.
\end{align}
As per focusing, note that since \(|C|/\sqrt{1+C^2} < 1\) for all real \(C\), a real focusing always exists.

\subsubsection{Case 2: $I = 0$}
In this case, solutions are: 
\begin{align}
\theta(t) &= \omega_c\tan\!\left[\omega_c\!\left(C - \frac{t}{2}\right)\right], \notag\\
\sigma_+(t) &= \sec^2\!\left[\omega_c\!\left(C - \frac{t}{2}\right)\right]\bigl[E\sin(\omega_c t) + F\cos(\omega_c t)\bigr], \notag\\
\sigma_\times(t) &= \sec^2\!\left[\omega_c\!\left(C - \frac{t}{2}\right)\right]\bigl[E\cos(\omega_c t) - F\sin(\omega_c t)\bigr], \notag\\
\omega(t) &= \frac{\omega_c}{2} + G\sec^2\!\left[\omega_c\!\left(C - \frac{t}{2}\right)\right],
\label{eq-34}
\end{align}
where
\begin{align}
C &= \frac{1}{\omega_c}\tan^{-1}\!\left(\frac{\theta_0}{\omega_c}\right), &
\{E,F,G\} &= \frac{\{\sigma_{+0},\;\sigma_{\times0},\;\omega_0 - \omega_c/2\}}{1 + (\theta_0/\omega_c)^2}.
\end{align}
These functions diverge at a time when
\[
t_f = 2C + \frac{(2n+1)\pi}{\omega_c}, \qquad n \in \mathbb{Z}.
\]
Thus, focusing occurs at finite time for the marginal case as well.

\subsubsection{Case 3: $I < 0$}
Solutions in this case mirrors the ones in the Case 1, with some juxtaposition:
\begin{align}
\theta(t) &= \frac{\omega_c C\cos(\omega_c t + D)}{\sqrt{1+C^2} + C\sin(\omega_c t + D)}, \notag\\
\sigma_+(t) &= \frac{E\sin(\omega_c t) + F\cos(\omega_c t)}{\sqrt{1+C^2} + C\sin(\omega_c t + D)}, \notag\\
\sigma_\times(t) &= \frac{E\cos(\omega_c t) - F\sin(\omega_c t)}{\sqrt{1+C^2} + C\sin(\omega_c t + D)}, \notag\\
\omega(t) &= \frac{\omega_c}{2} + \frac{G}{\sqrt{1+C^2} + C\sin(\omega_c t + D)},
\label{equ-36}
\end{align}
However with some different constants:
\begin{align}
C &= \sqrt{-\frac{I_0}{\omega_c^2}\left(1 - \frac{\theta_0^2}{4I_0} - \frac{\omega_c^2}{4I_0}\right)^2 - 1},\notag\\
D &= \tan^{-1}\!\left[\frac{2\sqrt{-I_0}}{\theta_0}\left(\frac{\omega_c}{2\sqrt{-I_0}} - \sqrt{1+C^2}\right)\right],\\
\{E,F,G\} &= \frac{\omega_c}{2\sqrt{-I_0}}\left\{\sigma_{+0},\;\sigma_{\times0},\;\omega_0 - \frac{\omega_c}{2}\right\}.
\end{align}
Note that no real solution exists for \(\sin(\omega_c t_f + D)\) in this case. Therefore, the denominator of ESR variables never vanish, and the trajectories never meet at a finite time.

These above solutions apply to both Region I and Region III because the equations of motion are identical in the absence of an electric field. However, the constants $(E,F,G)$ generally differ between the two regions due to the effect of the pulse in Region II. 

For $0 \le t \le T$, the central trajectory (starting from initial conditions $x_0, y_0, u_{x0}, u_{y0}$) is given by \cite{kar2024pulse}:
\begin{align}
x^{(II)}(t) &= \left(\frac{E_0}{B}\sin\alpha\right)t
+ \left(\frac{u_{x0}}{\omega_c} - \frac{E_0}{\omega_c B}\sin\alpha\right)\sin(\omega_c t) \notag\\
&\quad - \left(\frac{u_{y0}}{\omega_c} + \frac{E_0}{\omega_c B}\cos\alpha\right)\cos(\omega_c t)
+ \left(x_0 + \frac{u_{y0}}{\omega_c} + \frac{E_0}{\omega_c B}\cos\alpha\right), \label{eq:xII}\\[6pt]
y^{(II)}(t) &= -\left(\frac{E_0}{B}\cos\alpha\right)t
+ \left(\frac{u_{x0}}{\omega_c} - \frac{E_0}{\omega_c B}\sin\alpha\right)\cos(\omega_c t) \notag\\
&\quad + \left(\frac{u_{y0}}{\omega_c} + \frac{E_0}{\omega_c B}\cos\alpha\right)\sin(\omega_c t)
+ \left(y_0 - \frac{u_{x0}}{\omega_c} + \frac{E_0}{\omega_c B}\sin\alpha\right). \label{eq:yII}
\end{align}

To study focusing, consider a neighboring trajectory $x^{(II)\prime}, y^{(II)\prime}$ with infinitesimally different initial conditions. The linearized map from initial offsets $(\Delta x_0, \Delta y_0)$ to the separation at time $t$ is given by the Jacobian matrix $J(t)$. The condition for the two trajectories to meet (i.e., the separation to vanish) at a finite time $t_f$ is that $\det J(t_f) = 0$. This is evaluated using the matrix elements:

\begin{equation}
\begin{aligned}
J_{11}(t) &= \frac{\bigl(\frac{\theta_0}{2}+\sigma_{+0}\bigr)\sin(\omega_c t) - (\sigma_{\times0} - \omega_0)\cos(\omega_c t)}{\omega_c}
            + \frac{\sigma_{\times0} - \omega_0}{\omega_c} + 1, \\[6pt]
J_{12}(t) &= \frac{(\sigma_{\times0} + \omega_0)\sin(\omega_c t) - \bigl(\frac{\theta_0}{2}-\sigma_{+0}\bigr)\cos(\omega_c t)}{\omega_c}
            + \frac{\frac{\theta_0}{2}-\sigma_{+0}}{\omega_c}, \\[6pt]
J_{21}(t) &= \frac{\bigl(\frac{\theta_0}{2}+\sigma_{+0}\bigr)\cos(\omega_c t) + (\sigma_{\times0} - \omega_0)\sin(\omega_c t)}{\omega_c}
            - \frac{\frac{\theta_0}{2}+\sigma_{+0}}{\omega_c}, \\[6pt]
J_{22}(t) &= 1 + \frac{(\sigma_{\times0} + \omega_0)\cos(\omega_c t) + \bigl(\frac{\theta_0}{2}-\sigma_{+0}\bigr)\sin(\omega_c t)}{\omega_c}
            - \frac{\sigma_{\times0} + \omega_0}{\omega_c}.
            \end{aligned}
\end{equation}
The focusing condition is $\det \mathbb{J}(t_f) = J_{11}(t_f)J_{22}(t_f) - J_{12}(t_f)J_{21}(t_f) = 0$. The matrix elements involve the initial kinematic variables $\theta_0, \sigma_{+0}, \sigma_{\times0}, \omega_0$ evaluated at the start of Region II ($t=0^+$).

\subsection{Focusing Time Universality}

Remarkably, the evolution equations for the ESR variables in a pure magnetic field contain no explicit dependence on the electric field. The only role of the pulse is to set the initial conditions $(\theta_0,\sigma_{+0},\sigma_{\times0},\omega_0)$ at the beginning of Region II. Consequently, the focusing time $t_f$ for a congruence in Region II is given by the same formula as for a pure magnetic field \cite{shaikh2014kinematics}:
\begin{equation}
t_f = \frac{2}{\omega_c}\tan^{-1}\!\left(-\frac{\omega_c}{\theta_0 \pm 2\sqrt{I_0}}\right),
\end{equation}
where $I_0 = \sigma_{+0}^2 + \sigma_{\times0}^2 - \omega_0^2$. This universality means that, once the initial kinematic variables are fixed, the subsequent focusing behavior depends only on the magnetic field strength, not on the detailed history of how those initial conditions were produced. The electric pulse thus influences the focusing time only indirectly, through its effect on the initial values at the moment of injection.

\section{Memory-like effect probed via congruence}\label{section-IV}

\subsection{Imprint of memory as left in shear}

\begin{figure}[H]
    \centering

    \includegraphics[width=0.99\linewidth]{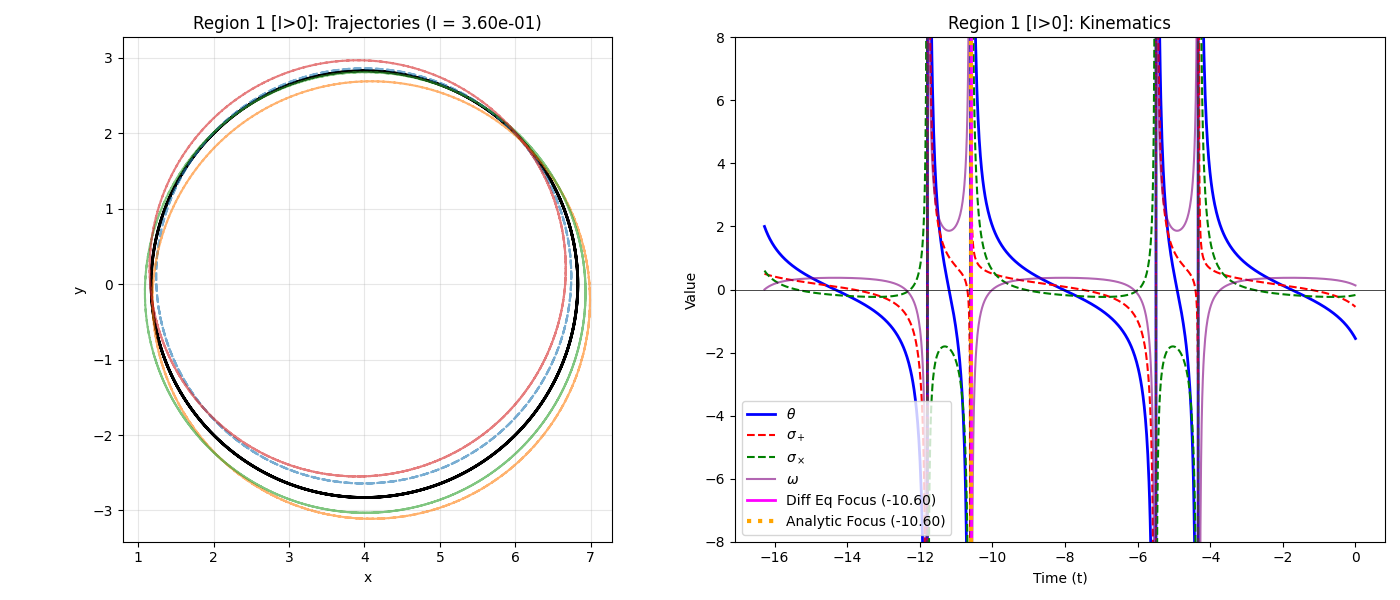}

    \par\vspace{0.1cm}
    {\small (a)}

    \vspace{0.1cm}

    \includegraphics[width=0.99\linewidth]{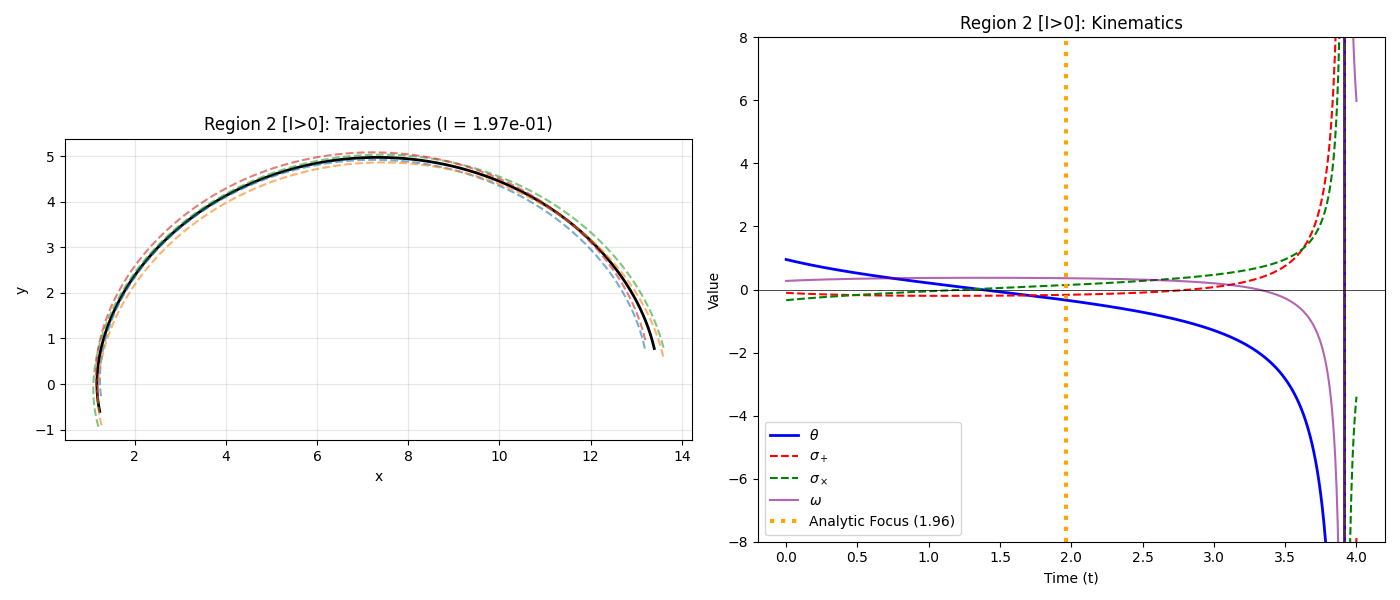}

    \par\vspace{0.1cm}
    {\small (b)}

    \vspace{0.1cm}

    \includegraphics[width=0.99\linewidth]{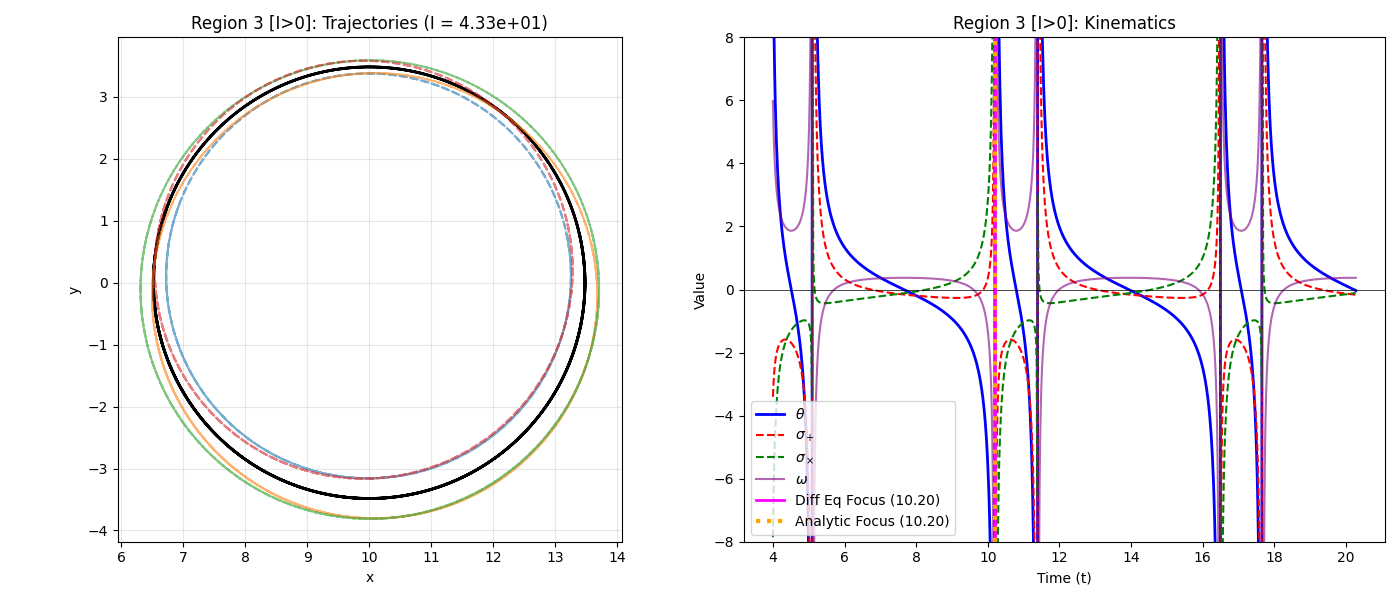}

    \par\vspace{0.1cm}
    {\small (c)}

    \caption{Trajectories and kinematics for $I>0$ in three regions: (a) Region I, (b) Region II, and (c) Region III.}
    \label{fig-4}
\end{figure}

The whole system is now broken down into three temporal regions based on the pulse passing: Region I (before the pulse), Region II (during the pulse), and Region III (after the pulse). In addition, we analyze them in the three regimes-$I<0$, $I=0$, $I>0$ separately. The trajectories and time evolution of associated ESR variables can be found in Figure \ref{fig-4} (shown for the $I>0$ regime as a representative case), in which we have plotted four different trajectories, marked by different colours corresponding to four different initial conditions. The central guiding path is highlighted in black in all the cases.

The evolution of the ESR variables across the three-region system is governed by the solutions mentioned before. These equations are solved analytically in closed form, with the solution in each region determined by the sign of the invariant $I$. The resulting expressions for $\theta$, $\sigma_+$, $\sigma_\times$, and $\omega$ are evaluated computationally over a discrete time array within each region, with the ESR state at the exit of one region passed directly as the initial condition to the next. All plots of the ESR variables are produced from these closed-form solutions.

\begin{figure}[H]
    \centering

    \includegraphics[width=0.6\linewidth]{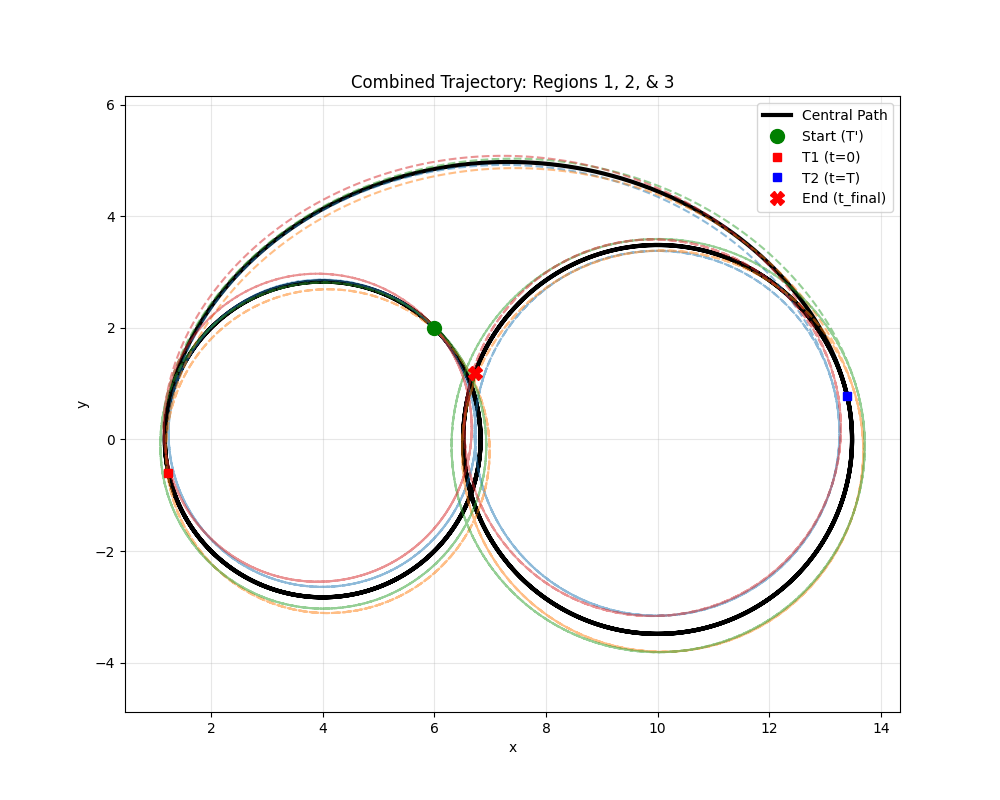}

    {\small (a)}

    \caption{Complete trajectory pictures: for $I>0$ . The position of the starting time, zero and $T$ (when the pulse hits and leaves) and the end time of simulation is marked on the trajectories. }
    \label{fig-7}
\end{figure}

Below these, in Figure \ref{fig-7}, one can find the complete trajectories combined in a single plot for $I>$ 0. Here it can be clearly seen that the memory of the pulse is stored in the radius of the cyclotron trajectories, as it jumps from once cycle to another as pulse passes, but further discussion shows that the ESR variables also do carry the memory of the pulse.

\begin{figure}[H]
    \centering
    \includegraphics[width=0.95\linewidth]{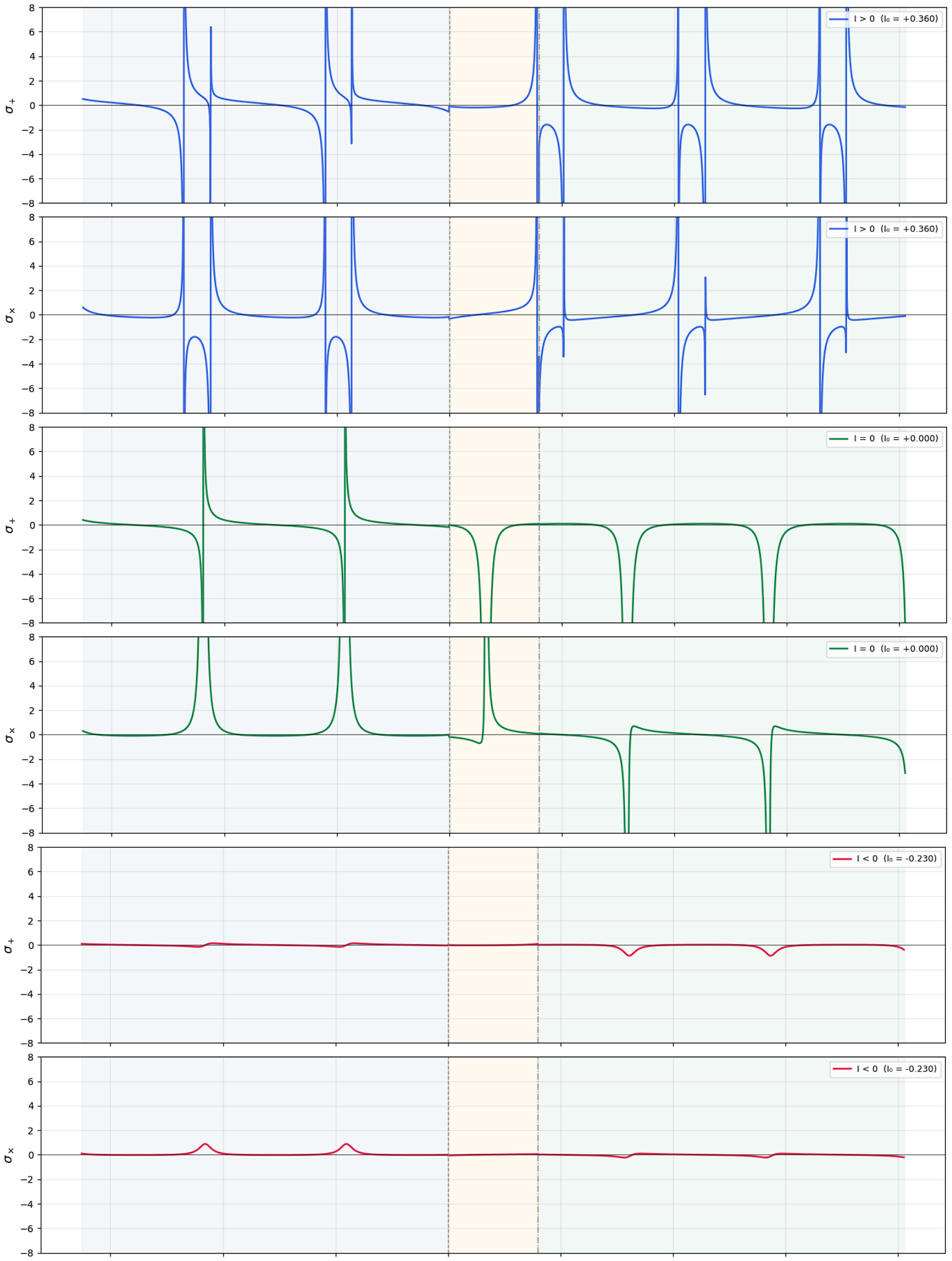}
    \caption{Evolution of shear components $\sigma_+$ and $\sigma_\times$ for all regimes ($I>0$, $I=0$, $I<0$), consolidated into a single figure for direct comparison. Regions 1, 2, and 3 are represented by blue, yellow, and green backgrounds respectively.}
    \label{fig-8}
\end{figure}

To understand this better, we have the combined shear evolution plot in Figure \ref{fig-8}. But what we really want to see is the change in nature of things from Region I to Region III, i.e. before and after the pulse. Now there is no noticeable change in expansion or rotation variables, as a result, there is no discernible change on the focusing behaviour as such. But there are interesting changes in the system's shear. Observing the shear only, across the three regions in Figure \ref{fig-8}, we can make an interesting observation,which we will discuss in the next subsection. Naively, there seems to be an interchange of the nature of the shear's evolution, between $\sigma\times$ and $\sigma+$, between regions I and III, with an additional sign flip. We will come back to this discussion in a later section.  

\subsection{Regression analysis of ESR shear variables: 
            verification of the analytic solution}

To rigorously verify that the numerically computed shear variables $\sigma_+$ and $\sigma_\times$ are consistent with the analytic ESR solution, we employ a rational-trigonometric regression model. Because this functional form maps directly to the analytic expressions for a charged particle in a pure magnetic field, it inherently provides the most accurate fit. The model represents the variables as a sinusoidal numerator over a sinusoidal denominator. If the computational simulation correctly captures the trajectory physics, this fit must yield $R^2 = 1.000$ \footnote{The coefficient of determination, denoted $R^2$, is a statistical measure that quantifies the goodness of fit of a regression model. For a set of $n$ data points $(y_i)$ with a mean $\bar{y}$ corresponding fitted values $(\hat{y}_i)$ obtained from the regression model, $R^2$ is given by:
\begin{equation}
R^2 = 1 - \frac{\displaystyle \sum_{i=1}^{n} (y_i - \hat{y}_i)^2}{\displaystyle \sum_{i=1}^{n} (y_i - \bar{y})^2}, \label{eq:R2}
\end{equation}}; any deviation serves as a precise indicator of a singularity (a focusing time) falling within the fitted window.

The fit is evaluated over exactly one period ($2\pi/\alpha \approx 6.283$ time units) ending at the boundary of each region, with time normalized to $\tau \in [0, 1]$. This deliberate isolation targets the settled, asymptotic behavior of the beam, allowing us to accurately extract the underlying analytical constants once initial transients have decayed.

\subsection*{Analytic Structure and Regression}

The analytic ESR solution in a pure magnetic field dictates the rational-trigonometric forms \ref{equ-31}, \ref{eq-34} ,\ref{equ-36}, depending on the value of I.  Because the regression model is structurally identical to these expressions, achieving an $R^2 = 1.000$ guarantees that the numerical procedure has successfully extracted the exact analytic constants ($C, D, E, F$). This serves as an independent, data-driven confirmation of the trajectory evolution.

\subsection*{$I < 0$: Quantifying the Classical Memory-Like Effect}

In the defocusing regime (we take $I_0 = -0.23$), the denominator remains strictly non-zero, meaning no focus occurs. Consequently, the rational-trigonometric model achieves a perfect $R^2 = 1.000$ across all variable--region combinations. The fitted numerator coefficients cleanly confirm the analytic quarter-period phase shift between $\sigma_+$ and $\sigma_\times$, with the sine and cosine roles exchanging seamlessly.

\begin{figure}[H]
    \centering
    \includegraphics[width=0.85\linewidth]{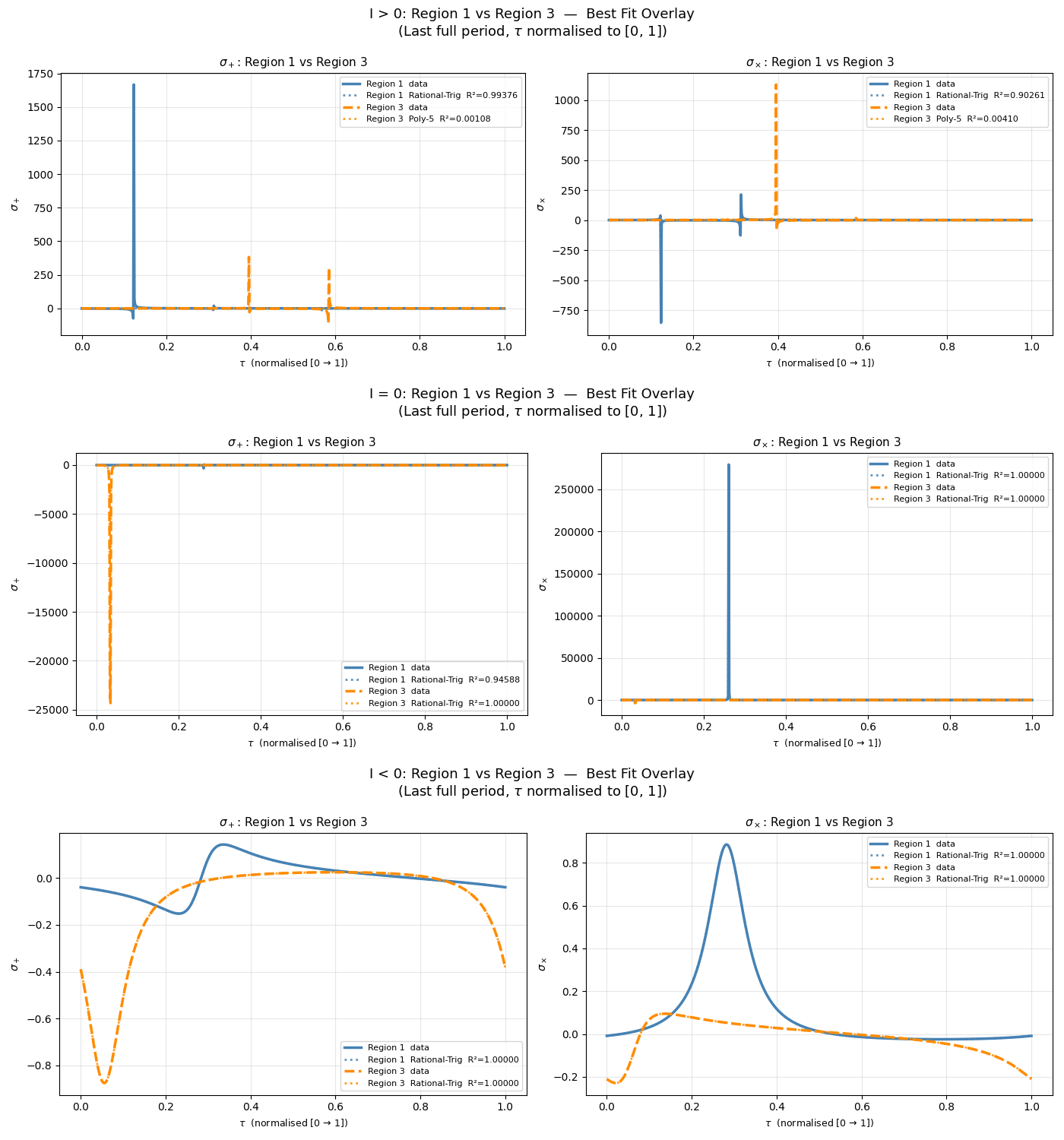}
    \caption{Comparison of $\sigma_+$ and $\sigma_\times$ in Regions I and III across all three invariant regimes ($I>0$, $I=0$, $I<0$). The panels demonstrate the quality of the rational-trigonometric regression fit, highlighting the exact parameter extraction in the defocusing ($I<0$) case and the singularities in the marginal ($I=0$) and focusing ($I>0$) regimes.}
    \label{fig:reg_combined}
\end{figure}

More importantly, the regression actively extracts the primary physical result of the simulation: a permanent shift in the denominator constants. In Region I, the pre-pulse denominator is governed by $C = -0.9273$ and $D = 0.1872$; post-pulse in Region III, these transition to $C = -0.3151$ and $D = -0.8920$. This shift directly quantifies the memory-like effect of the electric pulse. The drift velocity introduced during Region II permanently overwrites the beam's phase-space geometry, a structural change that the rational-trigonometric fit explicitly reads back from the cyclotron trajectories, which is visible as the fit and data, both completely overlap, as seen in Figure \ref{fig:reg_combined}.

The regression confirms the analytic solution in every regime where 
the extraction window is free of a focusing singularity. Beyond 
verification, the fitted constants provide a direct readout of the pulse's effect: comparing the denominator parameters 
between Region I and Region III reveals precisely how the pulse has 
altered the beam's phase-space structure. Where the fit fails, the 
failure is not a shortcoming of the method, it is a signal that the 
beam is approaching a focus, which is itself physically~meaningful.

\section{Complex Shear and Electromagnetic phase Memory }\label{section-V}

In this section, we try to quantify the idea of memory in the case we have been discussing so far. It is indeed clear that the memory-like effect is mostly based on the guiding centre of the cyclotron trajectory moving away on the plane, and somehow records its imprint in the shear component. But this needs to be elucidated further.

\subsection{Dynamical Phase}
We show here that the memory is encoded in a dynamical phase called the  $\theta_D$, which rotates the the $(\sigma_+, \sigma_\times)$ plane, and thus corresponds to a large gauge transformation relating the pre- and post-pulse vector potentials.
To show this, we define the \textbf{complex shear}:
\begin{equation}
\Sigma = \sigma_+ + i\sigma_\times.
\end{equation}
Which leads to:
\begin{align}
\frac{d\sigma_+}{dt} + i\frac{d\sigma_\times}{dt} + \theta(\sigma_+ + i\sigma_\times) &= \omega_c\sigma_\times - i\omega_c\sigma_+.
\end{align}
Note that $\sigma_\times - i\sigma_+ = -i(\sigma_+ + i\sigma_\times) = -i\Sigma$. Therefore:
\begin{equation}
\frac{d\Sigma}{dt} + \theta\Sigma = -i\omega_c\Sigma. 
\end{equation}
This is the fundamental evolution equation for the complex shear. For pure cyclotron motion, the expansion $\theta = 0$ (the area of a small square of particles remains constant). In that case:
\begin{equation}
\frac{d\Sigma}{dt} = -i\omega_c\Sigma
\implies 
\Sigma(t) = \Sigma(0) e^{-i\omega_c t}. 
\end{equation}
In this regime the magnitude $|\Sigma|$ is constant and the complex shear rotates uniformly at frequency $\omega_c$. Further $\sigma_+$ and $\sigma_\times$ have a fixed $\pi/2$ phase difference:
    \begin{equation}
    \sigma_+(t) = |\Sigma|\cos(\omega_c t - \phi_0), \qquad
    \sigma_\times(t) = -|\Sigma|\sin(\omega_c t - \phi_0).
    \end{equation}
This corroborates our earlier finding.
The analytic solutions for the ESR variables in a pure magnetic field, as written before, changes this and the complex shear takes the general form:
\begin{equation}
\Sigma(t) = \frac{Z e^{-i\omega_c t}}{\mathcal{D}(t)}, \label{eq:general_shear}
\end{equation}
where $Z = F + iE$ is a complex constant determined by the initial conditions, and $\mathcal{D}(t)$ is the denominator appropriate to the sign of the invariant $I$, as written in the table below.
\begin{table}[h!]
\centering
\begin{tabular}{|c|c|}
\hline
\textbf{Case} & \textbf{Denominator} $\mathcal{D}(t)$ \\
\hline
$I > 0$ & $C + \sqrt{1+C^2}\sin(\omega_c t + D)$ \\
$I = 0$ & $\cos^2\!\left[\omega_c\left(C - \frac{t}{2}\right)\right]$ \\
$I < 0$ & $\sqrt{1+C^2} + C\sin(\omega_c t + D)$ \\
\hline
\end{tabular}
\caption{The denominator $\mathcal{D}(t)$ for the three cases.}
\end{table}
Remember that the constants \(C\) and \(D\) are determined by the initial values of \(\theta_0\), \(\sigma_{+0}\), \(\sigma_{\times 0}\), and \(\bar{\omega}_0\).

During the pulse ($0 \le t \le T$), the electric field adds a drift motion. The guiding center moves according to:
\begin{equation}
\dot{X} = \frac{E_y}{B}, \qquad \dot{Y} = -\frac{E_x}{B}.
\end{equation}
For a square pulse ($E_x = E_0\cos\alpha$, $E_y = E_0\sin\alpha$, constant), we assume $\mathcal{E} = E_0 T$ is the integrated field strength. Then from the drift velocity $\vec{v}_{gc} = \dfrac{\vec{E} \times \vec{B}}{B^2}$, we can compute:
\begin{equation}
\Delta X = \frac{\mathcal{E}}{B}\sin\alpha, \qquad
\Delta Y = -\frac{\mathcal{E}}{B}\cos\alpha. 
\end{equation}
For our system, a standard result\footnote{The cyclotron orbit may be written as $\mathbf{r}(t)
=
\mathbf{R}_0
+
R(\cos\phi,\sin\phi)$.
A displacement of the guiding center
$\mathbf{R}_0 \to \mathbf{R}_0 + \Delta\mathbf{R}$
must be compensated by a shift of the angle variable
$\phi \to \phi + \theta_D$
to represent the same physical point.
Expanding to first order in $\theta_D$ and equating positions yields
$\Delta\mathbf{R}
+
R\,\theta_D(-\sin\phi,\cos\phi)
=0$.
Solving gives
\[
\theta_D
=
\frac{1}{R^2}
\left(
X_0\Delta Y - Y_0\Delta X
\right)
=
\frac{(\mathbf{R}_0 \times \Delta\mathbf{R})_z}{R^2}.
\]
Thus the dynamical phase equals the oriented area swept by the guiding center displacement, normalized by the orbit area scale $R^2$.} gives a dynamical kick, with the accumulated phase angle:
\begin{equation}
\theta_D = \frac{1}{R^2} \left( X_0 \Delta Y - Y_0 \Delta X \right). 
\end{equation}
This is the phase accumulated by the angle variable. Substituting:
\begin{equation}
\theta_D = -\frac{\mathcal{E}}{B R^2} \left( X_0 \cos\alpha + Y_0 \sin\alpha \right).
\end{equation}
This is the dynamical phase for our cyclotron memory system, which we can associate with the kick. Since the denominator $\mathcal{D}(t)$ is determined by the expansion and the invariant $I$, and since the pulse leaves $\mathcal{D}(t)$ unchanged up to a redefinition of constants, the only effect of the pulse is to rotate the numerator constant $Z = F + iE$, implying:
\begin{equation}
\Sigma_{\text{after}}(t) = \Sigma_{\text{before}}(t) \cdot e^{i\theta_D}. 
\end{equation}
This means the shear vector $(\sigma_+, \sigma_\times)$ rotates by $\theta_D$ in the $(\sigma_+, \sigma_\times)$ plane:
\begin{equation}
\begin{pmatrix}
\sigma_+^{\text{after}} \\[4pt]
\sigma_\times^{\text{after}}
\end{pmatrix}
=
\begin{pmatrix}
\cos\theta_D & -\sin\theta_D \\[4pt]
\sin\theta_D & \cos\theta_D
\end{pmatrix}
\begin{pmatrix}
\sigma_+^{\text{before}} \\[4pt]
\sigma_\times^{\text{before}}
\end{pmatrix}. 
\end{equation}
 Therefore, the constants $(E, F)$ themselves transform by the same rotation.
The effect of this rotation can be seen, e.g. in Figure \ref{fig-8} and in the regression analysis of the shear components.
\subsection{Connection to the Gauge Transformation}
As discussed neatly in \cite{kar2024pulse}, this kick in guiding center location of the cyclotron motion can be equivalently thought of as a gauge transformation.
The vector potential before and after the pulse is related by:
\begin{equation}
\vec{A}_{\text{after}} = \vec{A}_{\text{before}} + \nabla\Lambda,
\end{equation}
where
\begin{equation}
\Lambda(x,y) = -\mathcal{E} (x \cos\alpha + y \sin\alpha).
\end{equation}

This is a \textbf{large gauge transformation} because $\Lambda$ does not vanish at infinity ($\Lambda \to \pm\infty$ as $|x|,|y| \to \infty$). The dynamical phase angle is directly proportional to $\Lambda$ evaluated at the guiding center:
\begin{equation}
\theta_D = \frac{1}{B R^2} \Lambda(X_0, Y_0).
\end{equation}

\subsection{Memory in focusing times}
As we have discussed before, we can consider a congruence of two particles with the same orbit radius $R$ but different guiding center positions $(X_1, Y_1)$ and $(X_2, Y_2)$. Their orbital phases are:
\begin{equation}
\phi_1(t) = \omega_c t + \phi_1^{(0)}, \qquad
\phi_2(t) = \omega_c t + \phi_2^{(0)}.
\end{equation}
The focusing time $t_{\text{focus}}$ is the time when the two particles meet, i.e., when $\Delta\phi(t) = 0 \pmod{2\pi}$. For particles initially separated in phase:
\begin{equation}
t_{\text{focus}} = \frac{\phi_1^{(0)} - \phi_2^{(0)}}{\omega_c} \pmod{\frac{2\pi}{\omega_c}}.
\end{equation}
Now once the pulse passes, each particle acquires an additional phase shift as we have already worked out. The focusing time after the pulse is:
\begin{equation}
t_{\text{focus}}^{(\text{after})} = \frac{\Delta\phi_{\text{after}}}{\omega_c} = \frac{\Delta\phi_{\text{before}} + \left( \theta_D^{(2)} - \theta_D^{(1)} \right)}{\omega_c}.
\end{equation}
Substituting the expression for $\theta_D$, the change in focusing time before and after the pulse takes a nice structure:
\begin{equation}
    \Delta t_{\text{focus}} = -\frac{\mathcal{E}}{\omega_c B R^2} \left[ (X_2 - X_1) \cos\alpha + (Y_2 - Y_1) \sin\alpha \right].
\end{equation}
The change in focusing time $\Delta t_{\text{focus}}$ is \textbf{permanent}\footnote{The appearance of the negative sign in going from $\Delta(\Delta\phi)$ to $\Delta t_{\text{focus}}$ arises because the focusing time is determined by 
\emph{when} the phase difference returns to zero (modulo $2\pi$). An \emph{increase} in $\Delta\phi$ means the particles are now further apart in phase, 
requiring \emph{more time} to come back into alignment. Mathematically:
\begin{equation}
\Delta t_{\text{focus}} = -\frac{\Delta(\Delta\phi)}{\omega_c},
\end{equation}
where the minus sign converts a phase \emph{advance} (positive $\Delta(\Delta\phi)$) into a time \emph{delay} (positive $\Delta t_{\text{focus}}$).} — it remains after the pulse is gone. It is \textbf{proportional} to the integrated pulse strength $\mathcal{E} = \int E(t)\,dt$. This shift in focusing time is a direct, measurable signature of the electromagnetic memory effect. One can think, the focusing time shift is the electromagnetic analogue of the velocity memory effect in gravitational wave physics. 
In that context, a transient gravitational wave leaves freely falling test masses with a permanent velocity kick. 
Here, the cyclotron particles acquire a permanent \emph{phase} kick (equivalently, a shift in the orbital timing). However, a key difference is that gravitational memory manifests in \emph{real-space} separations, 
while electromagnetic cyclotron memory appears as a \emph{phase-space} (timing) effect.

\textcolor{black}{It is worth noting that while the present analysis utilizes a piecewise-constant electric field for analytical transparency, the underlying memory effect is entirely robust to the specific pulse profile. Studies of analogous gravitational memory have explicitly investigated continuous profiles (such as Gaussian or $\text{sech}^2$ pulses) and demonstrated that they yield permanent memory displacements consistent with those of simpler impulsive limits \cite{Zhang:2017-2,Zhang:2017-3,Chakraborty:2019yxn}. In our classical setup, the permanent shift in the dynamical phase, and consequently the complex shear, is fundamentally governed by the net guiding center displacement, which is determined entirely by the time-integral of the applied electric field (the integrated field strength $\mathcal{E}$ defined in Section IV). Therefore, a continuous profile, such as a Gaussian pulse, will yield qualitatively identical memory-like features provided the total integrated impulse remains the same.}

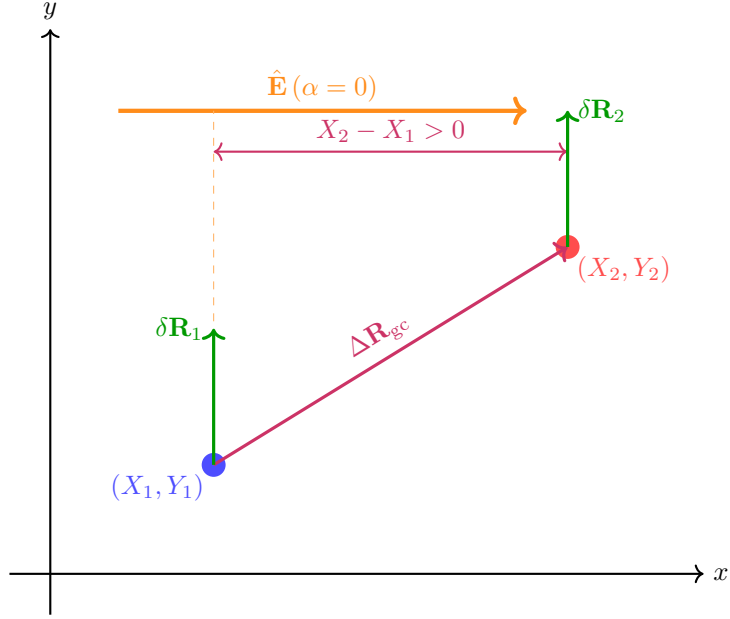
\begin{figure}[t]
\centering
\begin{tikzpicture}[scale=1.8]

    \draw[->, thick] (-0.3,0) -- (4.8,0) node[right] {$x$};
    \draw[->, thick] (0,-0.3) -- (0,4.0) node[above] {$y$};

    \coordinate (P1) at (1.2,0.8);
    \coordinate (P2) at (3.8,2.4);

    \fill[blue!70] (P1) circle (2.5pt);
    \fill[red!70]  (P2) circle (2.5pt);

    \node[blue!70, below left] at (P1) {$(X_1,Y_1)$};
    \node[red!70, below right] at (P2) {$(X_2,Y_2)$};

    \draw[->, very thick, purple!80]
        (P1) -- (P2)
        node[midway, above, sloped] {$\Delta \mathbf{R}_{\text{gc}}$};

    \draw[->, ultra thick, orange!90]
        (0.5,3.4) -- (3.5,3.4);
    \node[orange!90, above] at (2.0,3.4)
        {$\hat{\mathbf{E}}\,(\alpha=0)$};

    \draw[orange!60, dashed] (P1) -- (1.2,3.4);
    \draw[orange!60, dashed] (P2) -- (3.8,3.4);

    \draw[<->, purple!80, thick]
        (1.2,3.1) -- (3.8,3.1);
    \node[purple!80, above] at (2.5,3.1)
        {$X_2 - X_1 > 0$};

    \draw[->, very thick, green!60!black]
        (P1) -- ++(0,1.0)
        node[left] {$\delta\mathbf{R}_1$};

    \draw[->, very thick, green!60!black]
        (P2) -- ++(0,1.0)
        node[right] {$\delta\mathbf{R}_2$};


\end{tikzpicture}

\vspace{6pt}

\caption{
Geometric interpretation of the focusing time shift (shown for $\alpha=0$) for two particles with a shift in their guiding centers.
The differential phase shift depends on the projection
$\Delta \mathbf{R}_{\text{gc}} \cdot \hat{\mathbf{E}}$.
Here $X_2 > X_1$, so the projection is positive.
The dynamical phase difference is
$\theta_D^{(2)} - \theta_D^{(1)}
=
-\frac{\mathcal{E}}{BR^2}(X_2 - X_1)
< 0 .$
So,
the focusing time increases:
$\Delta t_{\text{focus}} > 0$
(delayed focusing).
}
\label{fig:focusing_geometry}
\end{figure}

\section{Conclusions}\label{section-VI}

Our work establishes the ESR formalism as a powerful diagnostic framework for detecting and characterizing memory-like effects in classical mechanical systems. By adopting a congruence-based perspective, we investigated the response of a pulse-driven charged particle in a uniform magnetic field through the evolution of the expansion, shear, and rotation variables. The ESR formalism provides a geometric description of the collective behaviour of neighboring trajectories and enables the identification of persistent imprints left by transient perturbations. Through an analysis of the congruence evolution and the associated focusing properties, we demonstrated that the passage of a finite electromagnetic pulse modifies the trajectory bundle in a manner that survives long after the pulse has ceased, thereby exhibiting a clear memory effect.

A central outcome of our analysis is the identification of the shear sector as the primary carrier of memory. While the expansion and rotation variables exhibit comparatively weaker sensitivity to the pulse history, the shear variables retain a pronounced and persistent record of the perturbation. This observation reveals that memory, when viewed from the perspective of congruence dynamics, is encoded in the geometric deformation of the trajectory bundle. To further assess the robustness and observability of this effect, we employed a logistic-regression-based analysis on synthetic trajectory data generated from the analytical solutions. The ability of the model to reliably distinguish and reconstruct the shear signatures demonstrates that the memory imprint can be inferred directly from trajectory information, thereby providing an independent validation of the analytical results and highlighting the experimental accessibility of the proposed framework. We should again stress that from an experimental perspective as well, the shear components $\sigma_+$ and $\sigma_\times$ map directly to measurable beam envelope observables: the beam's aspect ratio, orientation, and oscillation phase, where this memory-like effect can be observed.

Finally, by extending the discussion to complex shear and rotation variables, we uncovered a direct connection between congruence dynamics and electromagnetic phase memory. In this framework, the pulse-induced memory manifests itself through the phase structure of the complex ESR variables and can be understood in terms of residual gauge transformations generated by the transient perturbation. This connection provides a complementary geometric viewpoint of memory and reinforces the role of shear as the principal carrier of the information encoded by the pulse. Taken together, our results takes a formal step to establish the ESR formalism as a unified framework for understanding the interplay between trajectory deformation, shear memory, and gauge structures in classical dynamical systems. We hope to report more on these connections in future correspondences.

\section*{Acknowledgements}

The authors would like to thank Aritra Banerjee for suggesting the problem and his encouragement. 
MKD and MM acknowledge BITS Pilani for encouraging undergraduate research.

\section*{Declarations}

\subsection{Funding}

SD acknowledges
financial support from BITS Pilani in the form of a PhD fellowship.

\subsection{Conflict of Interest}

The authors declare that they have no conflict of interest.

\section*{Data Availability}
Data sharing is not applicable to this article, as no new datasets were generated or analyzed. All theoretical results and mathematical derivations necessary to reproduce the findings are fully contained in the manuscript.

\nocite{*}
\bibliographystyle{unsrt}
\bibliography{main}

\end{document}